\documentclass[reprint,aps,prb,superscriptaddress,amsmath,amssymb,bibnotes,showkeys]{revtex4-1}
\usepackage{graphicx}

\usepackage{amsmath}
\usepackage{dcolumn}
\usepackage{bm}
\usepackage{braket} 
\usepackage{bbold}
\usepackage{textcomp}
\usepackage{multirow}
\usepackage{url}
\usepackage{tabularx}
\usepackage{color}
\usepackage{soul}
\usepackage[
 colorlinks=true,
 urlcolor=blue,
 citecolor=blue,
 linkcolor=blue,
 bookmarks=false,
 pdfstartview={FitH},
]{hyperref}

\begin{document}

	\title{Native Defects in Antiferromagnetic Topological Insulator  MnBi$_2$Te$_4$}
	\author{Zengle~Huang}
	\affiliation{Department of Physics $\&$ Astronomy, Rutgers
		University, Piscataway, New Jersey 08854, United States}
	\author{Mao-Hua~Du}
	\affiliation{Materials Science and Technology Division, Oak Ridge National Laboratory, Oak Ridge, Tennessee 37831, United States}
	\author{Jiaqiang~Yan}
	\affiliation{Materials Science and Technology Division, Oak Ridge National Laboratory, Oak Ridge, Tennessee 37831, United States}
	\author{Weida~Wu}
	\email{wdwu@physics.rutgers.edu}
	\affiliation{Department of Physics $\&$ Astronomy, Rutgers
		University, Piscataway, New Jersey 08854, United States}
	
\begin{abstract}
Using scanning tunneling microscopy and spectroscopy, we visualized the native defects in antiferromagnetic topological insulator $\mathrm{MnBi_2Te_4}$. Two native defects $\mathrm{Mn_{Bi}}$ and $\mathrm{Bi_{Te}}$ antisites can be well resolved in the topographic images. $\mathrm{Mn_{Bi}}$ tend to suppress the density of states at conduction band edge. Spectroscopy imaging reveals a localized peak-like local density of state at $\sim80$~meV below the Fermi energy. A careful inspection of topographic and spectroscopic images, combined with density functional theory calculation, suggests this results from $\mathrm{Bi_{Mn}}$ antisites at Mn sites. The random distribution of $\mathrm{Mn_{Bi}}$ and  $\mathrm{Bi_{Mn}}$ antisites results in spatial fluctuation of local density of states near the Fermi level in $\mathrm{MnBi_2Te_4}$.   
\end{abstract}

\maketitle

The interplay between topology and magnetism in quantum materials is an active research front in condensed matter physics. The spontaneously broken time-reversal symmetry in magnetic topological insulator opens an exchange gap on the Dirac surface states~\cite{Qi2008,Tokura2019, Chen2010,Yu2010}. When the Fermi level is tuned into the exchange gap, interesting phenomena such as quantum anomalous Hall effect~(QAHE) could be realized\cite{Yu2010, Chang2013}. Previously QAHE was demonstrated in Cr- and V-doped (Bi,Sb)$_2$Te$_3$ thin films below 2~K~\cite{Chang2013,Chang2015,Mogi2015}, and lately in twisted bilayer graphene below 4~K~\cite{Serlin2020}. By engineering heterostructures of QAH films, another interesting topological quantum phase, axion insulator, has been realized in sandwich structure of magnetic topological insulators~\cite{Xiao2018,Mogi2017a,Mogi2017b}. These topologically protected quantum phases are promising platforms for fabricating high-speed and dissipationless electronics.
       
\begin{figure}[htpb]
	\includegraphics[width=\columnwidth]{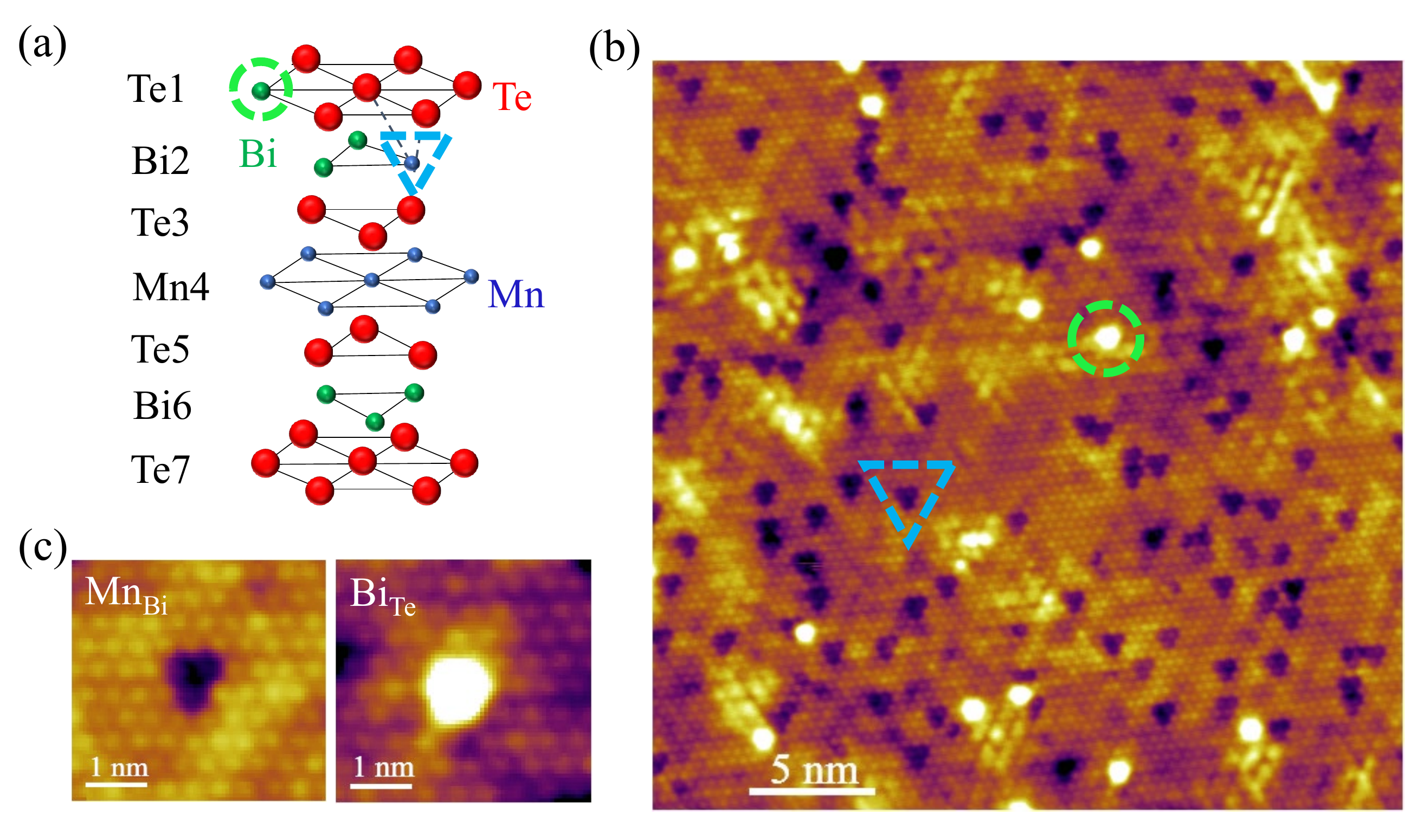}
	\caption{(Color online) (a) Definition of the atomic sites in the crystal structure of MnBi$_2$Te$_4$. (b) Topographic image of MnBi$_2$Te$_4$ ($30\times30$~ nm$^2$. Setpoint: $-0.8$~V, 0.3~nA). The blue triangle marks the $\mathrm{Mn_{Bi}}$ antisite at Bi2 site and the green circle marks a $\mathrm{Bi_{Te}}$ antisite at Te1 site. (c) Zoom-in images of a $\mathrm{Mn_{Bi}}$ antisite and a $\mathrm{Bi_{Te}}$ antisite.
\label{Fig1} 	}
\end{figure}

The recent prediction and discovery of antiferromagnetic topological insulator $\mathrm{MnBi_2Te_4}$ opens a new direction to achieve topologically protected quantum states in stoichiometric materials~\cite{Otrokov2019,Gong2019}. $\mathrm{MnBi_2Te_4}$ is a van der Waals compound comprised of Te-Bi-Te-Mn-Te-Bi-Te septuple layers. Each atomic plane within the septuple layer is a triangular lattice. The atomic planes are stacked in the ABC fashion. The structure can be viewed as intercalating an additional Mn-Te bilayer into the middle of topological insulator $\mathrm{Bi_2Te_3}$, as shown in FIG.~\ref{Fig1}(a). The magnetism comes from the Mn$^{2+}$ ions with high-spin configuration $S=\frac{5}{2}$ and magnetic moment of $\sim$5~$\mu_\textrm{B}$. Below N\'eel temperature $T_\textrm{N}$~$\approx$~25~K~\cite{Otrokov2019,Gong2019,Yan2019a,Lee2019}, magnetic moments of Mn$^{2+}$ within each septuple layer order ferromagnetically with an out-of-plane easy axis, and adjacent septuple layers couple antiferromagnetically, leading to an A-type antiferromagnetic structure~\cite{Yan2019a,Sass2020}. Theoretical calculations predict that $\mathrm{MnBi_2Te_4}$ can also host Weyl semimetal states in bulk, as well as QAH and axion insulating states in thin films~\cite{Otrokov2019b,Li2019a,Zhang2019}. Remarkably, by tuning Fermi level with backgate voltage, QAHE has been observed in $\mathrm{MnBi_2Te_4}$ flakes of odd-number septuple layers~\cite{Deng2020} and axion insulating state in flakes of even-number septuple layers~\cite{Liu2020a}. The QAHE persists up to 6.5~K, well above the quantization temperature of the best magnetically doped topological insulator thin films~\cite{Mogi2015}.

Despite the rapid progress in the realization of novel quantum states in $\mathrm{MnBi_2Te_4}$, some important materials issues remain unclear. For example, $\mathrm{MnBi_2Te_4}$ single crystals are $n$-type semiconductors with Fermi level about 200~meV above the conduction band minimum, as revealed by angle-resolved photoemission spectroscopy~(ARPES)~\cite{Hao2019,Li2019b,Chen2019,Otrokov2019,Gong2019,Lee2019} and tunneling spectroscopy~\cite{Yan2019a,Yuan2020}. As is well studied in other 3D topological insulators $\mathrm{Bi_2Te_3}$ and $\mathrm{Bi_2Se_3}$~\cite{Hashibon2011,Dai2016,West2012,Wang2013}, point defects can strongly affect the physical properties such as conductivity and magnetism by affecting the Fermi level. Yet, it is unclear which kind of defects causes the electron doping in $\mathrm{MnBi_2Te_4}$. X-ray diffraction indicates antisite defects in both Mn and Bi sites, and possibly Mn vacancies~\cite{Zeugner2019}. Previous scanning tunneling microscopy (STM) found $\mathrm{Mn_{Bi}}$ antisites~\cite{Yan2019a,Yuan2020}, without in-depth investigation on how they affect the electronic structure. Nanoscale local density of states~(LDOS) fluctuations near Fermi level is observed by scanning tunneling spectroscopy (STS) and could be related to defect-induced local fluctuation of Bi and Te orbitals~\cite{Yuan2020}. When the Fermi level is inside the bulk bands because of doping by native defects, the charge transport would be dominated by bulk carriers. An high backgate voltage~\cite{Deng2020} is needed to tune the Fermi level to observe fascinating quantum states, which is challenging for device application. Therefore, it is imperative to identify and understand the native defects in topological materials like $\mathrm{MnBi_2Te_4}$ for better control of defect species and concentrations.
   
In this paper, we present STM/STS studies on the native defects in antiferromagnetic topological insulator $\mathrm{MnBi_2Te_4}$.  Our results reveal $\sim 3 \%$ of Mn$_\textrm{Bi}$ antisites and $\sim 0.2 \%$ of Bi$_\textrm{Te}$ antisites in nominal $\mathrm{MnBi_2Te_4}$. Mn$_\textrm{Bi}$ defects suppress the LDOS at the conduction band edge.  The spectroscopy mapping reveals that the LDOS at $E_\textrm{F}$ correlates with the local density of Mn$_\textrm{Bi}$ defects. In addition, We observe a significant peak-like LDOS at $\sim80$~meV below $E_\textrm{F}$ in localized triangular regions. DOS calculated by density functional theory~(DFT) indicates this LDOS peak comes from $\mathrm{Bi_{Mn}}$ antisites. These findings suggest important roles of the native defects in affecting the electronic structure and Fermi level DOS of $\mathrm{MnBi_2Te_4}$, which is critical for the observation of topological phenomena like QAHE. The identification of point defects could facilitate the optimization of $\mathrm{MnBi_2Te_4}$ synthesis through defect tuning. 

$\mathrm{MnBi_2Te_4}$ single crystals were grown by the flux method described in Ref.~\onlinecite{Yan2019a}. STM/STS measurements were performed at 4.5~K in an Omicron LT-STM with base pressure 1$\times$10$^{11}$~mbar. Electrochemically etched tungsten tips were characterized on a clean Au~(111) surface before STM experiments. $\mathrm{MnBi_2Te_4}$ single crystals were cleaved \textit{in situ} at room temperature and immediately inserted into the cold STM head. Scanning tunneling spectroscopy measurements were performed with standard lock-in technique with modulation frequency 455~Hz and amplitude 10~mV.

All calculations are based on DFT~\cite{Hohenberg1964,Kohn1965} implemented in the VASP code~\cite{Kresse1996}. The interaction between ions and electrons is described by the projector augmented wave method~\cite{Kresse1999}. The Perdew-Burke-Eznerhof (PBE) exchange correlation functional~\cite{Perdew1996} and a kinetic energy cutoff of 270~eV were used. A $U$ parameter of 4~eV was applied to Mn 3$d$ orbitals~\cite{Dudarey1998} and the DFT-D3 vdW functional~\cite{Grimme2010} was used to account for the weak interlayer interaction, following several previous DFT studies.~\cite{Li2019a,Chen2019,Liu2020a} Lattice parameters of  $\mathrm{MnBi_2Te_4}$ were optimized, and atomic positions were relaxed until the forces are less than 0.02 eV/\AA{}. The optimized lattice parameters are a = 4.365~\AA{} and c = 40.476~\AA{}, respectively, in good agreement with the experimentally measured values of 4.3338 \AA{} and 40.931~\AA{}~\cite{Yan2019a,Yan2019b}. A 3$\times$3$\times$2 supercell (54 formula units in six septuple layers) and a 2$\times$2$\times$1 $k$-point mesh were used for the calculation of the $\mathrm{Bi_{Mn}}$ defect in $\mathrm{MnBi_2Te_4}$. The c axis was doubled to allow the A-type antiferromagnetic ordering. The DOS of the defect-containing supercell was calculated using a denser 4$\times$4$\times$1 $k$-point mesh.

\begin{figure}[htpb]
	\includegraphics[width=\columnwidth]{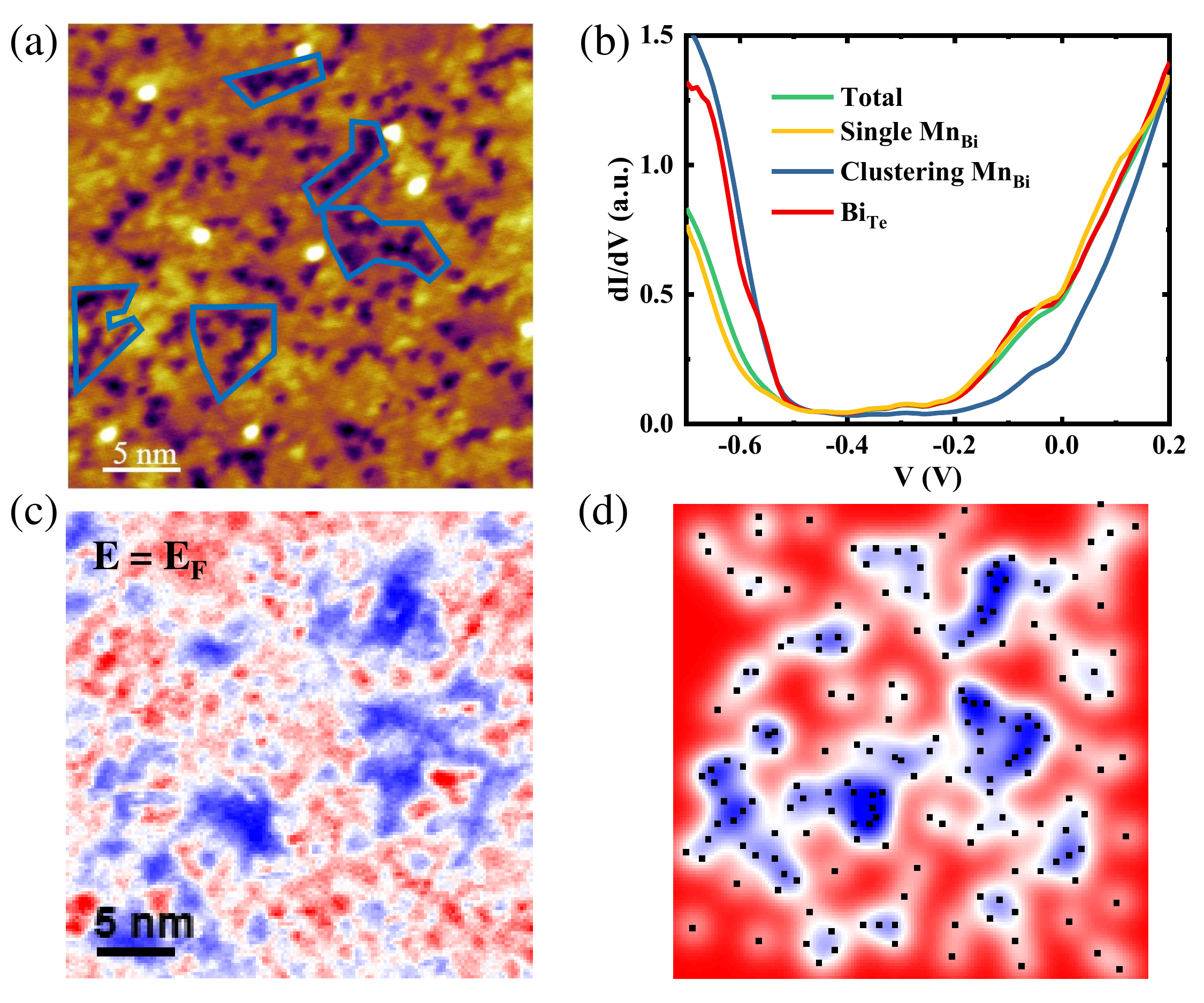}
	\caption{\label{Fig2}  (Color online) (a) STM topographic image of $\mathrm{MnBi_2Te_4}$. Blue dashed lines mark the region of clustering $\mathrm{Mn_{Bi}}$ antisites ($30\times30$~nm$^2$. Setpoint: $-$1~V, 0.5~nA). (b) Spatially averaged tunneling spectra obtained from the field of view in FIG.~3(a) (Setpoint: 0.4~V, 1~nA). The green curve is the averaged spectra of the whole area; the yellow and red curves are the averaged spectra of isolated $\mathrm{Mn_{Bi}}$ and $\mathrm{Bi_{Te}}$, respectively; the blue curve is the averaged spectra of the clustering MnBi antisites marked by blue dashed lines in (a). (c) Local conductance ($dI/dV$) mapping at $E_\textrm{F}$ showing spatial local density of state fluctuations. (d) Simulated influence map of $\mathrm{Mn_{Bi}}$ antisites using Gaussian function. The black dots denote the locations of the $\mathrm{Mn_{Bi}}$ antisites.  
	}
\end{figure}

\begin{figure*}[htpb]
	\includegraphics[width=2\columnwidth]{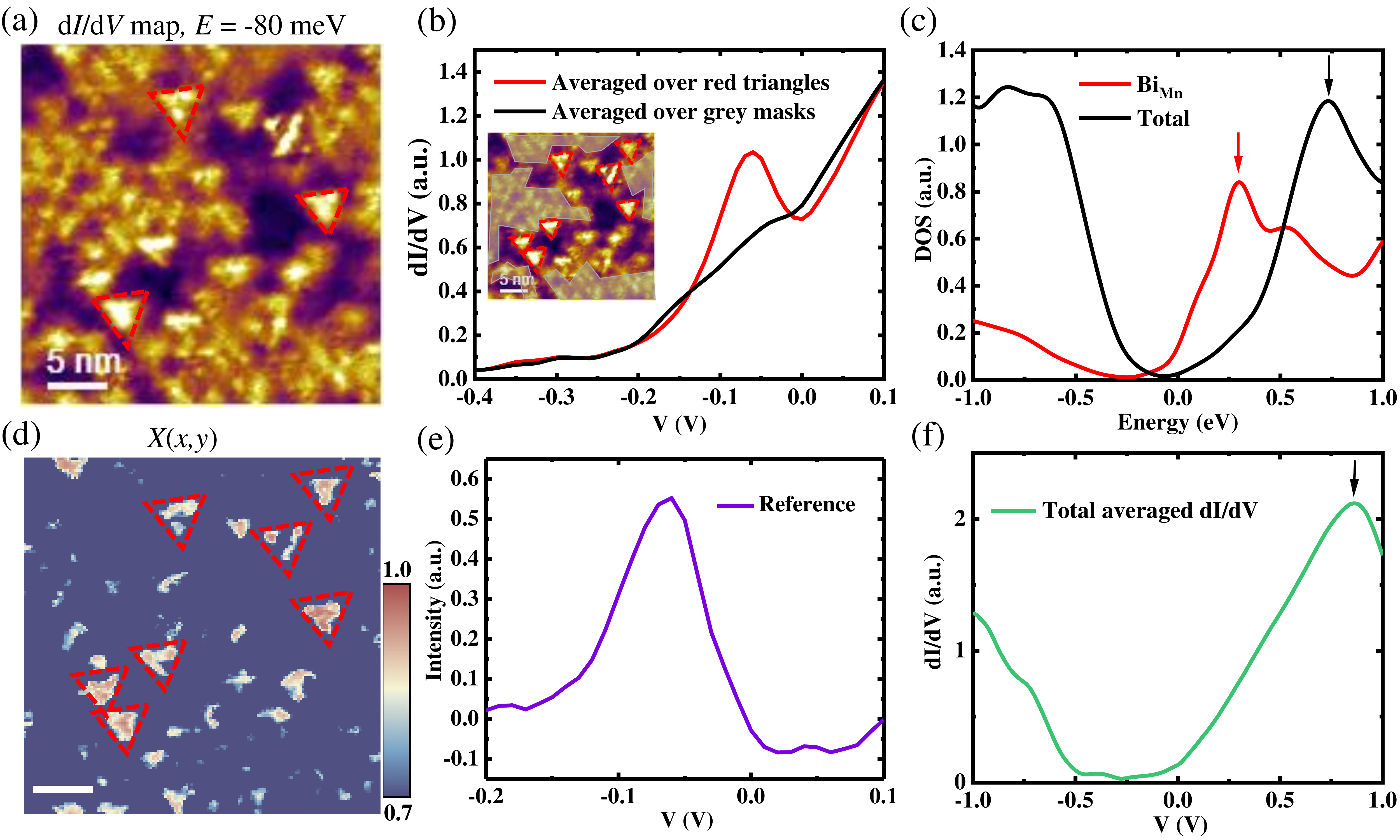}
	\caption{ (Color online) (a) Local conductance mapping at $-$80~meV in the same field of view as FIG.~\ref{Fig2}(a). Red dashed triangles highlights bright triangular regions of high LDOS at $-$80~meV. (b) Tunneling spectra of the different bright areas in (a) (Set point: 0.2~V, 1~nA). The red curve is the averaged tunneling spectrum of bright triangles. The black curve is the averaged tunneling spectrum of the rest bright areas where the peak around $-$80~meV is absent. The area used for averaging is shown in the inset. (c) DFT calculation of DOS of a $\mathrm{MnBi_2Te_4}$ supercell containing 54 formula units. The black curve is the total DOS of the whole supercell while the red curve is the partial DOS of a $\mathrm{Bi_{Mn}}$ defect in the supercell. Total DOS is normalized to a similar scale as the partial DOS of a $\mathrm{Bi_{Mn}}$ defect. The energy of the conduction band minimum is set to zero. The red arrow indicates the DOS peak of a $\mathrm{Bi_{Mn}}$ antisite and the black arrow indicates a peak in total DOS about 0.7~eV above conduction band minimum. (d) The spectrum correlation map $X(x, y)$, where the value of each pixel is the cross-correlation coefficient between the $dI/dV$ spectrum (background removed) and the reference in (e).  The red dashed triangles mark the same defects in the inset of (b). (e) Reference spectrum with background removed. 
		\label{Fig3}	}
\end{figure*}
A representative STM topographic image of cleaved (001) surface of $\mathrm{MnBi_2Te_4}$ is shown in Figure~\ref{Fig1}(b). Like previous STM topograhpic images~\cite{Yan2019a,Yuan2020}, two types of defects can be recognized: the dark triangular defects marked by the blue triangle and the bright protrusive ones by the green circle. Atomic resolution allows clear identification of the defect position, as shown in Fig.~\ref{Fig1}(c). The dark triangular defect centers at Bi atoms in the second layer, while the bright defect centers at the topmost Te atoms. Similar dark triangular defects have been observed in the previous STM work on Mn-doped Bi$_2$Te$_3$ and are identified as Mn atoms substituting the Bi atoms in the second layer~($\mathrm{Mn_{Bi}}$ antisite)~\cite{Hor2010}. The bright defects have also been seen in various 3D topological insulators such as Bi$_2$Se$_3$ and Sb$_2$Te$_3$, and are identified as the pnictogen atoms substituting the topmost chalcogen atoms~\cite{Dai2016,Jiang2012}. Thus we assign the dark triangular defects to $\mathrm{Mn_{Bi}}$ antisites in the second layer and the bright defects to $\mathrm{Bi_{Te}}$ antisites in the topmost layer. The density of $\mathrm{Mn_{Bi}}$ is $(3.0\pm0.1)\%$ and that of $\mathrm{Bi_{Te}}$ is about $(0.17\pm0.02)\%$. In Mn-doped Bi$_2$Te$_3$ merely $2\%$ of doped Mn is sufficient to induce long-range ferromagnetic order~\cite{Lee2014}. Even though there is no experimental evidence for additional magnetic contribution associated with $\mathrm{Mn_{Bi}}$, a recent work suggests antiferromagnetic alignment between the moments of $\mathrm{Mn_{Bi}}$ antisites and Mn layer~\cite{Liu2020b}. Apart from $\mathrm{Mn_{Bi}}$ and $\mathrm{Bi_{Te}}$, there are other unknown defect-like features in FIG.~\ref{Fig1}(b). It is difficult to identify them because they overlap with $\mathrm{Mn_{Bi}}$ and $\mathrm{Bi_{Te}}$. Samples with low defect-density are desirable for identifying unknown native defects.

To explore the influence of $\mathrm{Mn_{Bi}}$ and $\mathrm{Bi_{Te}}$ on the electronic structure of $\mathrm{MnBi_2Te_4}$, $dI/dV$ mapping is performed in the field of view of Fig.~\ref{Fig2}(a). From the $dI/dV$ map we could obtain spatially averaged tunneling spectra for different areas on the surface. Fig.~\ref{Fig2}(b) presents the tunneling spectra averaged over isolated $\mathrm{Mn_{Bi}}$, $\mathrm{Bi_{Te}}$ and the total area. Note that several $\mathrm{Mn_{Bi}}$ antisites are  closed to each others in some locations as highlighted by blue dashed lines in Fig.~\ref{Fig2}(a). The ``clustering'' of $\mathrm{Mn_{Bi}}$ antisites is due to spatial fluctuation of random distribution. The averaged tunneling spectra of these areas are labeled as ``clustering $\mathrm{Mn_{Bi}}$'' in Fig.~\ref{Fig2}(b). Consistent with the previous STM results~\cite{Yuan2020}, no significant spectroscopic feature associated with defect states was observed on either $\mathrm{Mn_{Bi}}$ or $\mathrm{Bi_{Te}}$ from $-0.7$~V to 0.2~V.

However, in the areas of clustering $\mathrm{Mn_{Bi}}$, the LDOS above conduction band minimum ($\sim-0.2$~eV) is significantly suppressed. This suppression is also observed in the $dI/dV$ map at $E_\textrm{F}$ in Fig.~\ref{Fig2}(c), where regions with reduced LDOS correlate with $\mathrm{Mn_{Bi}}$ antisites in Fig.\ref{Fig2}(a). To quantify the correlation between LDOS fluctuation at $E_\textrm{F}$ and $\mathrm{Mn_{Bi}}$ antisites, we extract the positions of $\mathrm{Mn_{Bi}}$ defects and compute the cross-correlation coefficient between the defect influence map in Fig.~\ref{Fig2}(d) and the  $dI/dV$ map in Fig.~\ref{Fig2}(c). The defect influence is modeled by a Gaussian function~\cite{Dai2016}. The cross-correlation coefficient reaches the maximum value of 0.46 when the influence radius is $\sim 1$~nm, which is comparable with the apparent size of $\mathrm{Mn_{Bi}}$. The substantial positive correlation corroborates the suppression of the LDOS at conduction band bottom by  $\mathrm{Mn_{Bi}}$.  This suppression of conduction band states can be explained by the fact that the conduction band in  $\mathrm{MnBi_2Te_4}$ is dominated by Bi $p$-orbitals~\cite{Otrokov2019}, and thus the lack of Bi in the $\mathrm{Mn_{Bi}}$-dense area reduces the LDOS in the conduction band. In addition, $\mathrm{Mn_{Bi}}$ antisites are also acceptors as previously reported in Mn-doped Bi$_2$Te$_3$~\cite{Hor2010} and they are supposed to shift the Fermi level towards the bulk band gap. 

Besides direct visualization of defects in the topographic images, point defects can also be revealed by spectroscopic mapping, especially for the defects with distinct electronic fingerprints such as a LDOS peak at their characteristic energy. An unknown kind of defects deep in the septuple layer was revealed in the $dI/dV$ map at $-80$~meV shown in Fig.~\ref{Fig3}(a). This $dI/dV$ map is at the same field of view as in Fig.~\ref{Fig2}(a). There are several bright triangles highlighted by the red dashed lines, indicating the LDOS at $-80$~meV is enhanced in these areas. The averaged tunneling spectra are shown in Fig.~\ref{Fig3}(b) and the areas used for averaging are marked in the inset. Red curve is averaged over bright triangles and clearly, there is a pronounced peak centered at $-80$~meV. In comparison, although there are other non-triangular bright areas in Fig.~\ref{Fig3}(a), the pronounced peak is absent in their averaged tunneling spectra represented by the black curve in Fig.~\ref{Fig3}(b). These areas also light up because of the fluctuation of conduction band DOS due to the $\mathrm{Mn_{Bi}}$ mentioned above. 

To understand the origin of the LDOS peak, we compute the partial DOS of a $\mathrm{Bi_{Mn}}$ antisite in a $\mathrm{MnBi_2Te_4}$ supercell with 54 formula units using DFT. The results are shown in Fig~\ref{Fig3}(c). The overall shape of the DFT-calculated total DOS agrees with the averaged tunneling spectrum in Fig~\ref{Fig3}(f). For the tunneling spectrum, there is a broad peak about 1~eV above CBM indicated by a black arrow. It resembles the peak in DFT-calculated total DOS which is 0.7~eV above CBM as indicated by a black arrow in Fig.~\ref{Fig3}(c). The  $\mathrm{Bi_{Mn}}$ defect induces a partial DOS peak (of Bi-6$p$ character) at about 0.3~eV above the conduction band minimum (CBM), indicated by a red arrow. Compared with DFT calculation, the experimental LDOS peak is about 0.1~eV closer to the CBM. The shift might originates from the different environment in STM measurements, where the defect position is close to the surface~\cite{Dai2016}. The agreement between tunneling spectroscopy measurement and DFT calculation suggests that the LDOS peak may come from $\mathrm{Bi_{Mn}}$ antisite. On the other hand, other potential defects like Mn vacancy and $\mathrm{Bi_{Te}}$ deep in the septuple later are expected to perturb mainly the valence band states. 

\begin{figure}[htbp]
	\includegraphics[width=\columnwidth]{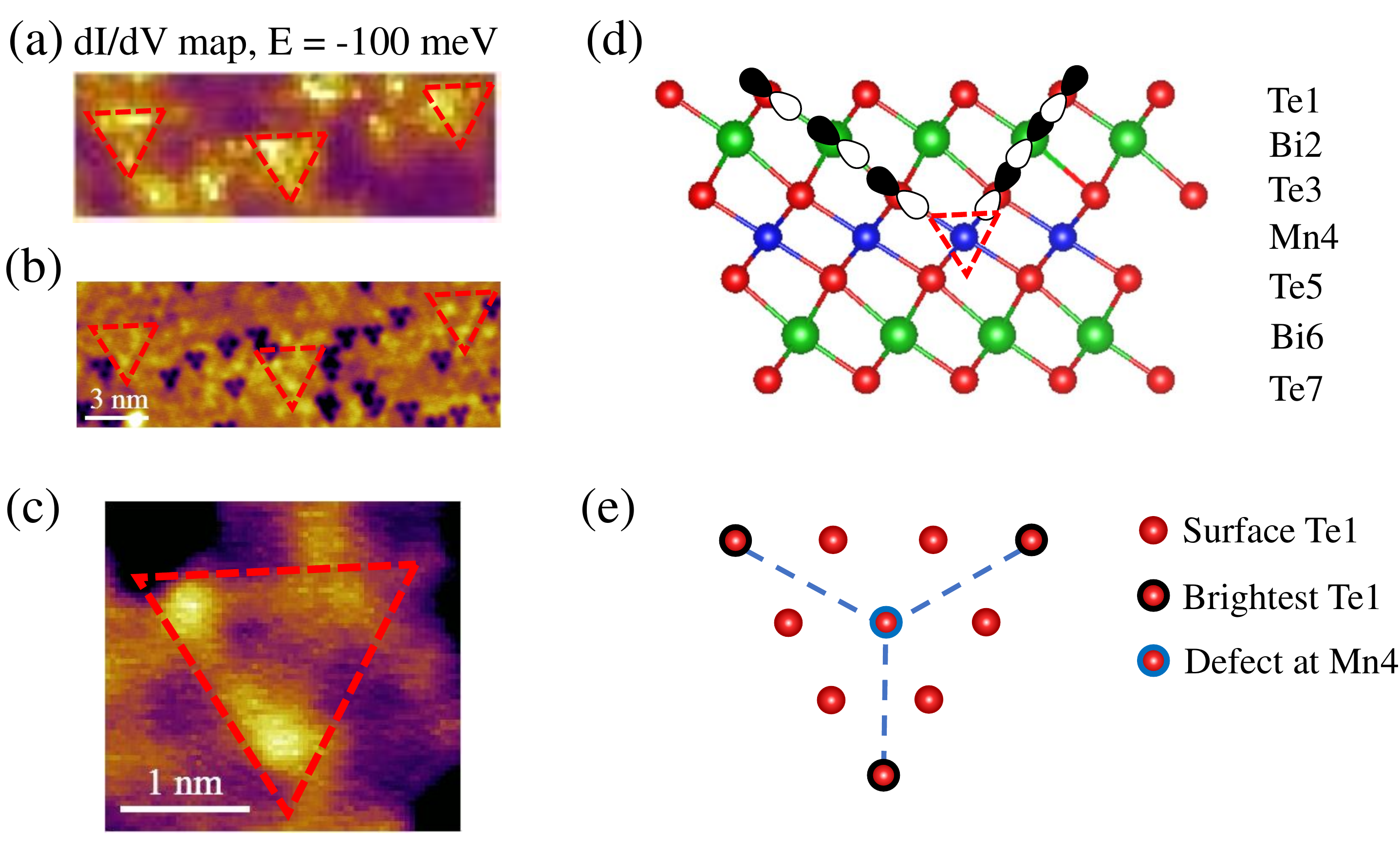}
	\caption{ (Color online) (a) Local conductance mapping at $-$100 meV. The red dashed triangles mark the triangular regions with pronounced peak around $-$100 meV in the tunneling spectra. (b) Topographic image of the same area in (a) ($20\times20$~nm$^2$. Set point: 1~V, 0.5~nA). (c) The zoom-in topographic image of $\mathrm{Bi_{Mn}}$ as marked by red dashed triangle in (b). (d) Illustration of $pp\sigma$ chains in $\mathrm{MnBi_2Te_4}$ in the presence of a defect at Mn4 site. The black and white lobes are the p orbitals of Bi and Te atoms, and the d orbital of the Mn atom. (e) Schematics of a defect at Mn4 plane whose defect state propagates to the topmost layer along the $pp\sigma$ chains, brightening three Te atoms.  
		\label{Fig4} 	}
\end{figure}

To better highlight the bright triangles with the LDOS peak at $-80$~meV, we compute cross correlation $X(x, y)$ between the spectrum at each pixel and a reference spectrum with the LDOS peak (red curve in Fig.~\ref{Fig3}(b)). The averaged spectrum~(green curve in Fig.~\ref{Fig2}(b)) was subtracted from the spectrum at each pixel $(x,y)$ and the reference spectrum. The resultant spectra are denoted as $G(x, y)$ and  $R$ (shown in Fig.~\ref{Fig3}(e)), respectively. The cross-correlation $X(x, y)$ is computed using the following equation: 
\begin{equation}
X(x,y) = 
\frac{\overline{G(x, y)\cdot R}-\overline{G(x, y)}\cdot\overline{R}}{{\sigma_G\cdot\sigma_{R}}}
\label{e:crosscorr}
\end{equation}
where $\overline{G(x, y)}$ and $\overline{R}$ are the mean,  $\sigma_G^2$ and $\sigma_{R}^2$ are the variance for $G(x, y)$ and reference respectively. The computation is performed in the interval of $-0.2$ to $0.1$~V, where the peak is centered. The map of $X(x, y)$ is shown in Fig.~\ref{Fig3}(d). Most bright triangles in the $dI/dV$ map are preserved and they are distinctly highlighted.

The similar peak feature was reported in a recent STM study and was attributed to some defect-induced local fluctuation of Bi and Te orbitals~\cite{Yuan2020}. Since the LDOS peak is closed to Fermi energy, it could affect the transport properties and also the local magnetism via itinerant exchange, which are important for the realization of QAH and axion insulating states.  Our spectroscopy map and spectrum correlation map $X(x, y)$ reveal that the LDOS peak only appears in localized triangular regions, which means it comes from localized electronic states of the defects deep in the septuple layer. 

To determine the position of the defect with LDOS peak in the septuple layer, we correlate the triangular regions in spectroscopic maps with topographic images. $dI/dV$ maps are measured simultaneously with the topographic images, so we overlay the $dI/dV$ map in Fig.~\ref{Fig4}(a) on the topography in Fig.~\ref{Fig4}(b) and obtain the approximate positions of the defects in topography. As shown in Fig.~\ref{Fig4}(b) and (c), there are three bright dots forming a triangle on top of the triangles in $dI/dV$ maps. The distance between the dots is about 1.3~nm, which is approximately three times the lattice constant ($\sim 4.32$~\AA)~\cite{Yan2019a}. The bonding scheme of $\mathrm{MnBi_2Te_4}$ is similar to that of other 3D topological insulators like  $\mathrm{Bi_2Se_3}$ and $\mathrm{Sb_2Te_3}$. Each atoms form six $\sigma$ bonds with its closest neighbors in the two adjacent atomic planes with atomic $p$ orbitals (or $d$ orbitals for Mn atoms). The influence of a defect deep in the septuple layer could propagate to the topmost Te plane along the three $pp\sigma$ chains passing through the defect atom~\cite{Urazhdin2002,Jiang2012}, as shown in Fig.~\ref{Fig4}(d). Since STM is most sensitive to the topmost Te atoms, the three surface Te atoms terminating the chains would appear as the most prominent features, as shown in Fig.~\ref{Fig4}(e)~\cite{Dai2016}. According to the $pp\sigma$ bonding argument, the size of defect suggests that it is located at the Mn4 plane. Prior X-ray diffraction refinement suggests the plausible existence of $\mathrm{Bi_{Mn}}$ antisites and Mn vacancies~\cite{Zeugner2019}, both of which are at the Mn4 plane. Combining the defect position analysis with STS measurements and DFT calculation, we believe this defect is a $\mathrm{Bi_{Mn}}$ antisite instead of a Mn vacancy. This is consistent with the defect formation energy calculation which favors $\mathrm{Bi_{Mn}}$ over Mn vacancies~\cite{Du2020}.

In summary, our STM/STS studies reveal three kinds of native defects: $\mathrm{Mn_{Bi}}$, $\mathrm{Bi_{Te}}$, and $\mathrm{Bi_{Mn}}$ in single crystals of antiferromagnetic topological insulator $\mathrm{MnBi_2Te_4}$.  $\mathrm{Mn_{Bi}}$ significantly suppresses the LDOS at conduction band edge, and $\mathrm{Bi_{Mn}}$ possesses a localized electronic state around 80~meV below $E_\textrm{F}$, contributing to a pronounced peak in the LDOS. These findings stress the importance of precise control of native defects and thus Fermi level in $\mathrm{MnBi_2Te_4}$, which is crucial for observing topological phenomena related to topological surface states or chiral edge states. In addition, high density of defects in current crystals prevent the unambiguous identification of defects deep in the septuple layer. Low-defect-density single crystals are desired for further STM investigation of native defects in $\mathrm{MnBi_2Te_4}$.                  

The STM work at Rutgers was supported by NSF Grants No.~DMR-1506618 and No.~EFMA-1542798. W.W. acknowledges the support from an ARO grant No.~W911NF2010108. The crystal growth and DFT calculation at ORNL was supported by the US Department of Energy, Office of Science, Basic Energy Sciences, Materials Sciences and Engineering Division.

This manuscript has been co-authored by UT-Battelle, LLC under Contract No.~DE-AC05-00OR22725 with the U.S. Department of Energy. The United States Government retains and the publisher, by accepting the article for publication, acknowledges that the United States Government retains a non-exclusive, paid-up, irrevocable, world-wide license to publish or reproduce the published form of this manuscript, or allow others to do so, for United States Government purposes. The Department of Energy will provide public access to these results of federally sponsored research in accordance with the DOE Public Access Plan (http://energy.gov/downloads/doe-public-access-plan).

\bibliography{MBTdefect}
\end{document}